\documentclass{article}
\usepackage{natbib}
\usepackage{graphics}
\usepackage{graphicx}
\usepackage{latexsym}
\usepackage{a4}
\usepackage{a4wide}
\begin{document}

\hspace*{10.0cm}\parbox{4.0cm}{
OCU-HEP 2002-01 \\
}

\begin{center}
\begin{Large}
\begin{bf}
AFM pictures of the surfaces of glass RPC electrodes  
damaged by water vapor contamination
\end{bf}
\end{Large}
\end{center}

\begin{center}
T.~Kubo, E.~Nakano and Y.~Teramoto
\end{center}

\begin{center}
\begin{it}
Institute for Cosmic Ray Physics, Osaka City University,
Osaka 558-8585, Japan
\end{it}
\end{center}

\begin{abstract}
We present surface pictures of the damaged electrodes from the 
Glass Resistive Plate Chambers (GRPCs) taken by an 
Atomic Force Microscope (AFM). 
For the test, a set of chambers were operated with freon mixed gas (damaged)
and freonless gas (not damaged), contaminated with 1000 $\sim$ 2000 ppm of 
water vapor. In the AFM pictures, clear differences in damage are seen 
between the electrodes in the chambers with the freon mixed gas and 
the freonless gas; a combination of freon and water vapor caused the damage.
\end{abstract}


\input{epsf}
The reason for the permanent efficiency drop in the GRPCs
operated (in the streamer mode) with the chamber gas, contaminated 
with 1000 $\sim$ 2000 ppm of water vapor, has been studied 
since its finding 
during the development \cite{teramoto95,teramoto2000,teramoto2001} 
of the Belle detector at KEKB. In this letter, we report 
Atomic Force Microscope (AFM) pictures of the surfaces of 
electrodes from the chambers that were used for the damage test
that was previously reported \cite{teramoto2001}. The samples were cut
from the 2 mm thick glass electrodes.
In the test, an efficiency drop was observed 
if the chambers were operated with 
a freon mixed gas (Ar/C$_{4}$H$_{10}$/CH$_{2}$FCF$_{3}$ = 25/25/50),
but no significant drop was seen when using a freonless 
mixture (Ar/C$_{4}$H$_{10}$) with the same level of water 
contamination (1000 $\sim$ 2000 ppm) in the gas. 

\begin{figure}
\input epsf
\begin{center}
\leavevmode
\epsfxsize=6.0cm
\epsfbox{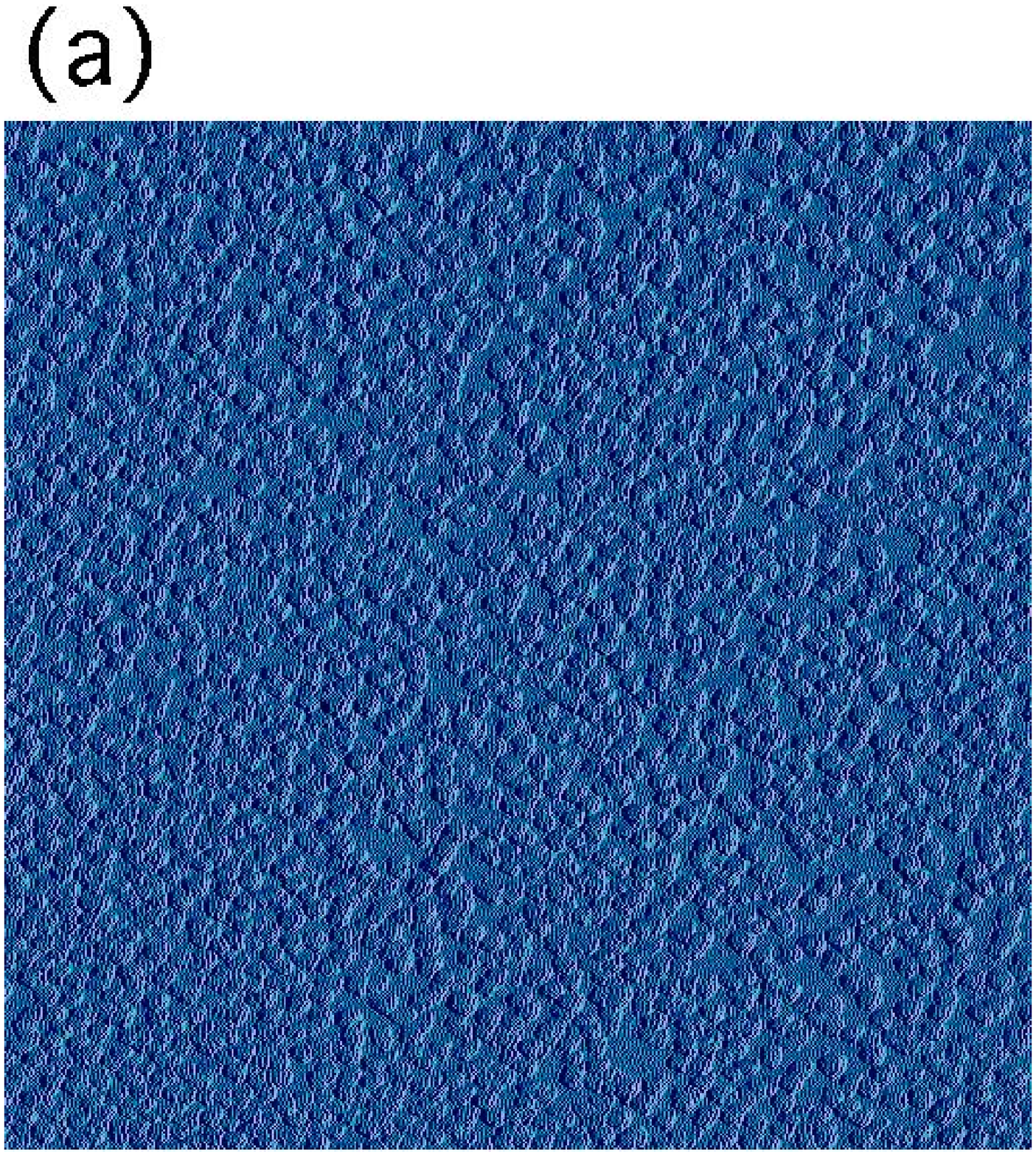}
\epsfxsize=6.0cm
\hspace*{0.5cm}\epsfbox{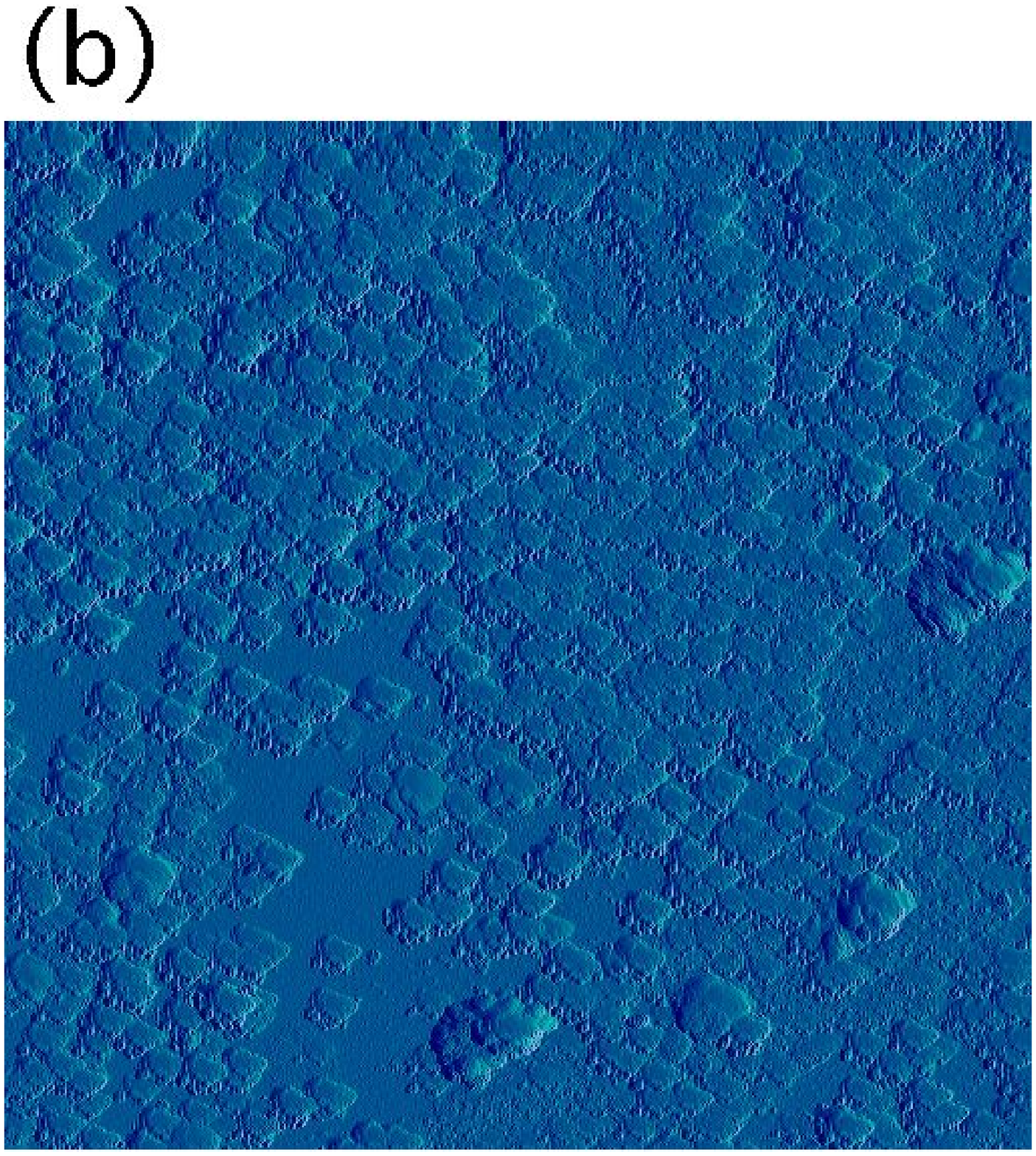}
\newline
\epsfxsize=6.0cm
\epsfbox{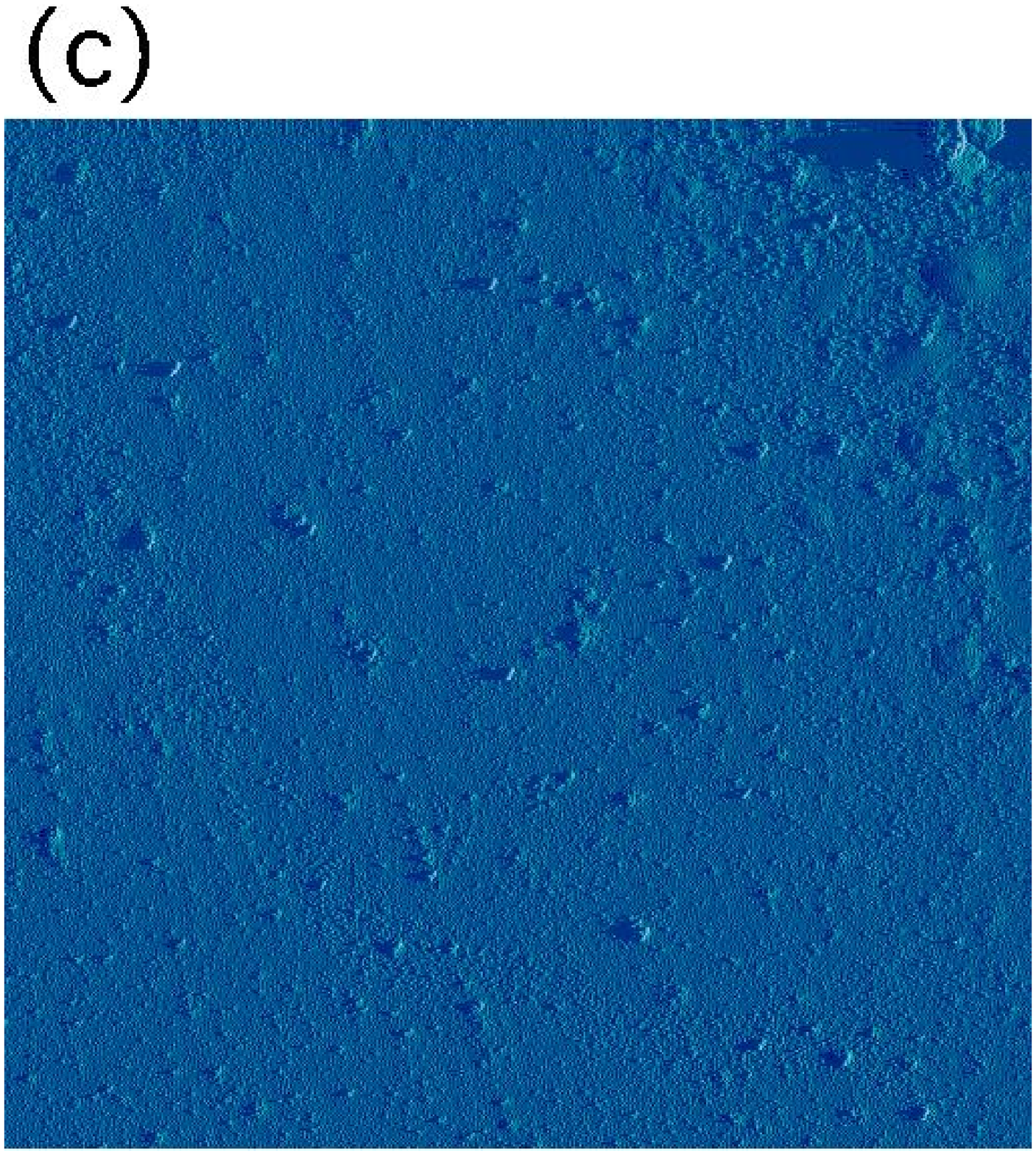}
\epsfxsize=6.0cm
\hspace*{0.5cm}\epsfbox{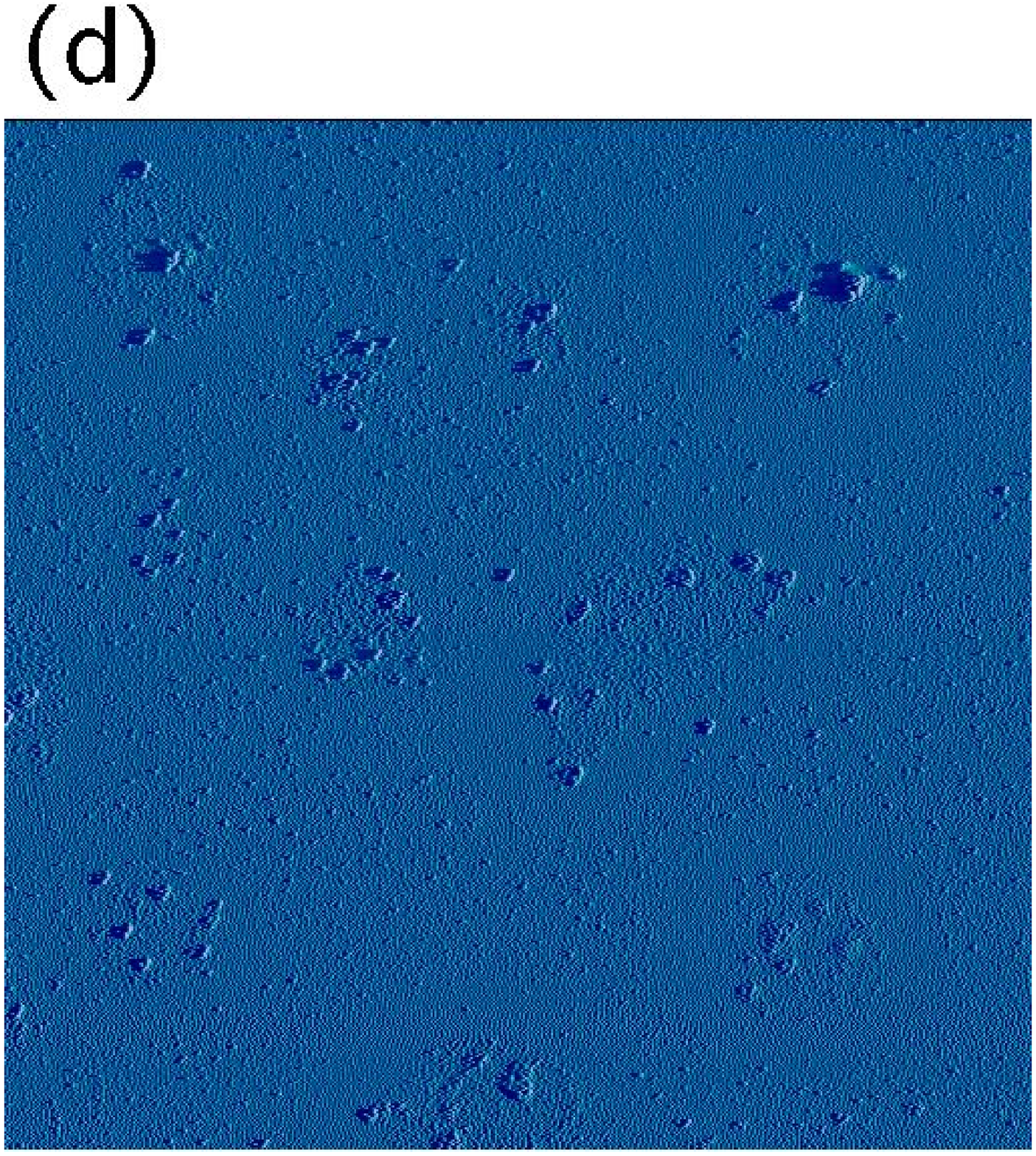}
\newline
\epsfxsize=6.0cm
\epsfbox{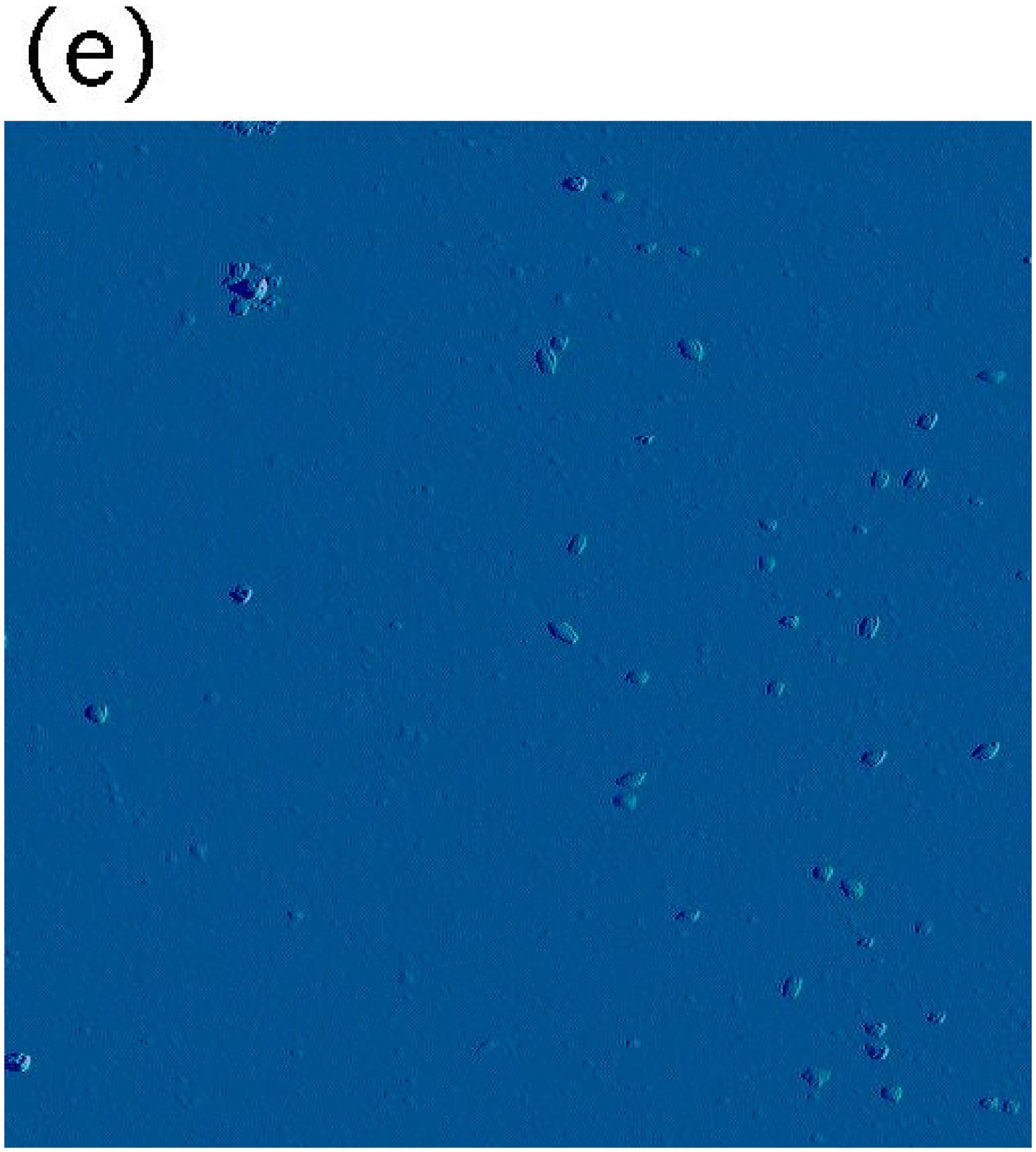}
\caption{AFM pictures of glass surfaces. (a) anode (freon),
(b) cathode (freon), (c) anode (freonless), (d) cathode (freonless),
(e) brand new glass. The size of each frame is 10 $\times$ 10 $\mu$m.}
\label{fig:AFM}
\end{center}
\end{figure}

The AFM pictures of the glass surfaces are shown in Fig.~\ref{fig:AFM}. 
In the freon operated chamber, substances of 0.05 $\mu$m in diameter 
and 100 nm thickness are seen on the anode. 
On the cathode, larger diameter (0.08 $\mu$m) and irregular shaped substances 
are seen. 
We consider the substances as deposits on the glass based on these pictures.
This explanation is also supported by the fact that the damaged
electrodes can be recovered by scrubbing them using paper dipped in
alcohol \cite{teramoto2000}.
In the freonless chamber, the observed substances are much 
smaller (0.005 $\mu$m), both on the anode and cathode. 

In conclusion, the AFM picture clarified the scenario of GRPC
damage; when glass RPCs are operated in the streamer mode 
in a chamber gas containing freon and operated under water 
vapor-contaminated conditions, deposits are formed 
on both the anode and the cathode surfaces. 
These substances reduce the work-function of the electrode surfaces, 
leading to the field emission of electrons from the cathode.
This reduces the electric field inside the chamber below the  
efficiency plateau and causes a permanent efficiency drop.

\noindent{\em Acknowledgments.}
\smallskip \\
\hspace*{12pt}The authors thank Dr. C. Lu for suggesting us to examine the 
surfaces of the damaged glass surface by AFM. 
We also thank all of the members 
of the KLM subdetector group in the Belle collaboration for their 
many useful discussions and information about glass RPCs.


\end{document}